\documentclass[english]{article}
\usepackage[utf8]{luainputenc}
\usepackage{color}
\usepackage{amsbsy}
\usepackage{esint}

\makeatletter
\usepackage{fullpage}

\makeatother

\usepackage{babel}
\begin{document}

\title{Dynamical spacetime symmetry}

\author{Benjamin C. Lovelady\thanks{Utah State University Dept of Physics, email: benjamin.lovelady@aggiemail.usu.edu}
\ and James T. Wheeler\thanks{Utah State University Dept of Physics, email: jim.wheeler@usu.edu}}
\maketitle
\begin{abstract}
According to the Coleman-Mandula theorem, any gauge theory of gravity
combined with an internal symmetry based on a Lie group must take
the form of a direct product in order to be consistent with basic
assumptions of quantum field theory. However, we show that an alternative
gauging of a \emph{simple} group can lead \emph{dynamically} to a
spacetime with compact internal symmetry. The biconformal gauging
of the conformal symmetry of n-dimensional Euclidean space doubles
the dimension to give a symplectic manifold. Examining one of the
Lagrangian submanifolds in the flat case, we find that in addition
to the expected SO(n) connection and curvature, the solder form necessarily
becomes Lorentzian. General coordinate invariance gives rise to an
SO(n-1,1) connection on the spacetime. The principal fiber bundle
character of the original SO(n) guarantees that the two symmetries
enter as a direct product, in agreement with the Coleman-Mandula theorem.
\end{abstract}

\section{Introduction}

The Coleman-Mandula theorem \cite{ColemanMandula} and generalizations
\cite{Percacci} show that, given certain assumptions likely true
of a satisfactory quantum field theory, any unification of general
relativity with internal symmetries based on Lie groups must take
the form of a direct product of the Poincaré or conformal group with
a compact internal symmetry group. By extending to a graded Lie algebra,
supersymmetric theories escape this conclusion and give a non-trivial
unification of gravity with the standard model interactions. In the
present examination, we find an alternative to such unification, using
a quotient of the conformal group of Euclidean space which doubles
the original dimension to a symplectic manifold. We show that the
solution of the field equations \emph{dynamically} produces a Lorentzian
metric on a Lagrangian submanifold. The class of orthonormal frame
fields of this Lorentzian metric is invariant under $SO\left(n-1,1\right)$
and enters as a direct product with the $SO\left(n\right)$ symmetry
of the original principal fiber bundle, satisfying the Coleman-Mandula
theorem. Of course, while this approach satisfies the Coleman-Mandula
theorem without supersymmetry, it does not preclude supersymmetric
extension.

Our method uses the standard construction of a Cartan geometry \cite{Kobayashi},
in which a principal fiber bundle is produced from the quotient of
a Lie group by a Lie subgroup. The Lie subgroup provides the local
symmetry over the quotient manifold. Group quotients allow group-theoretic
insights into gravity models, and thus improve on the early work of
Utiyama \cite{Utiyama} and Kibble \cite{Kibble1}, who extended global
Lorentz and Poincaré symmetries, respectively, to local symmetries.
The group quotient method was first employed for gravity by Ne'eman
and Regge \cite{Neeman,ReggeN} who used it as a systematic way to
study general relativity and supergravity. Later, Ivanov and Niederle
used the same method to develop a variety of gauge theories, including
the biconformal gauging we use here \cite{IvanovI,Ivanov}, though
the appropriate gravity field equations appear to arise from the curvature-linear
action found by Wehner and Wheeler \cite{WW}.

Starting from the generators of the conformal group of Euclidean $n$-space,
we present the biconformal gauging (developed in \cite{Ivanov,WW,NCG}
and presented concisely in \cite{Hazboun Wheeler}). Rather than continuing
by introducing Cartan curvatures, we work directly with the homogeneous
quotient manifold, using a known solution to the flat Maurer-Cartan
equations. This solution is naturally expressed in terms of a pair
of involute $1$-forms which together span a 2n-dimensional symplectic
manifold. The Killing metric induces a Lorentzian metric on one of
the resulting Lagrangian submanifolds (\cite{Spencer Wheeler}, also
see \cite{Hazboun Wheeler,Hazboun Wheeler-1}), and we interpret this
submanifold as spacetime.

In previous work studying these solutions \cite{Hazboun Wheeler,Hazboun Wheeler-1},
the $SO\left(4\right)$ connection was separated into an $SO\left(3,1\right)$
piece and additional fields. The additional fields, constructed from
certain invariant scalar fields, were interpreted as contributions
to dark matter and dark energy. The $SO\left(3,1\right)$ piece of
their decomposition becomes compatible with the usual soldering of
the base manifold to the fiber.

The novelty of our approach is to leave the $SO\left(4\right)$ symmetry
intact, and distinct from the Lorentzian symmetry that necessarily
\cite{Spencer Wheeler} develops for the metric. Because the gauging
includes the $SO\left(n\right)$ symmetry of the original Euclidean
space, the fiber symmetry includes its own $SO\left(n\right)$ connection.
However, the solder form now fails to ``solder'', displaying instead
a Lorentzian inner product and inducing the spacetime metric. General
coordinate covariance induces local Lorentz symmetry on orthonormal
frame fields, so the final model has both $SO\left(n\right)$ and
$SO\left(n-1,1\right)$ symmetries and corresponding connections,
in the direct product form required by the Coleman-Mandula theorem.
We introduce adapted coordinates that enable us to clearly display
the simultaneous presence of the two symmetries.

In a further new contribution to understanding these spaces, we include
a discussion of the meaning of the full biconformal manifold. We show
that the homogeneous solution permits identification of the full $2n$-dimensional
space as the co-tangent bundle of spacetime with orthogonal Lagrangian
submanifolds, one of which may be made flat by conformal transformation.

While we deal with only the flat case here, we expect further examination
to permit curvatures of both connections. For the $n=4$ case, the
$SO\left(4\right)$ symmetry is naturally replaced by $SU\left(2\right)\times SU\left(2\right)$,
and interpreted as a left-right symmetric electroweak model. Symmetry
breaking of one of the $SU\left(2\right)$ groups should then give
a grav-electroweak unification. Larger $n$ could incorporate $SO\left(n\right)$
or $Spin\left(n\right)$ GUT models.

\section{Biconformal Gauging}

\subsection{Quotient manifold method}

To develop the biconformal space, we use the quotient manifold method
\cite{Kobayashi,Neeman,ReggeN}. Starting with a Lie group $\mathcal{G}$,
(in our case the conformal group, $\mathcal{C}$), we construct a
principal fiber bundle by taking the quotient by a Lie subgroup, $\mathcal{H}$
(in our case the Euclidean Weyl group, $\mathcal{W}$, comprised of
$SO\left(n\right)$ transformations and dilatations). This subgroup
becomes the local symmetry of the $2n$-dimensional quotient manifold.

Lie groups have natural connection 1-forms $\boldsymbol{\omega}^{A}$
dual to the group generators $G_{A}$, defined by the linear mapping
$\boldsymbol{\omega}^{A}\left(G_{B}\right)=\delta_{B}^{A}$, with
the coordinate bases being dual, $\left\langle \frac{\partial}{\partial x^{\mu}},\mathbf{d}x^{\nu}\right\rangle =\delta_{\mu}^{\nu}$.
Rewriting the commutation relations of the generators
\begin{equation}
\left[G_{A},G_{B}\right]=c_{AB}^{\qquad C}G_{C},\label{eq:StructConst}
\end{equation}
where $c_{\;BC}^{A}$ are the group structure constants, in terms
of these dual $1$-forms, we arrive at the Maurer-Cartan structure
equations, 
\begin{equation}
\mathbf{d}\boldsymbol{\omega}^{A}=-\frac{1}{2}c_{\;BC}^{A}\boldsymbol{\omega}^{B}\wedge\boldsymbol{\omega}^{C}\label{eq:LieStructure}
\end{equation}
The integrability conditions of eq.(\ref{eq:LieStructure}) follow
from the Poincare lemma, $\mathbf{d}^{2}=0$, and exactly reproduce
the Jacobi identity,
\[
\left[G_{A},\left[G_{B},G_{C}\right]\right]+\left[G_{B},\left[G_{C},G_{A}\right]\right]+\left[G_{C},\left[G_{A},G_{B}\right]\right]=0
\]
so that eq.(\ref{eq:LieStructure}) is equivalent to the Lie algebra.

It is convenient to define $\mathbf{C}_{\;B}^{A}\equiv c_{\;CB}^{A}\boldsymbol{\omega}^{C}$
so the structure equation becomes $\mathbf{d}\boldsymbol{\omega}^{A}=\frac{1}{2}\boldsymbol{\omega}^{C}\wedge\mathbf{C}_{\;C}^{A}$.

The quotient of $\mathcal{G}$ by a Lie subgroup divides the Maurer-Cartan
into horizontal and vertical parts. Thus, if we write the connection
forms as $\tilde{\boldsymbol{\omega}}^{A}=\left(\tilde{\boldsymbol{\omega}}_{\mathcal{H}}^{a},\tilde{\boldsymbol{\omega}}_{\mathcal{M}}^{m}\right)$
with $\tilde{\boldsymbol{\omega}}_{\mathcal{H}}^{a},a,b,c=1,\ldots K$
spanning $\mathcal{H}$ and $\tilde{\boldsymbol{\omega}}_{\mathcal{M}}^{m},m,n,p=1,\ldots N-K$
spanning the homogeneous quotient manifold, $\mathcal{M}^{N-K}=\mathcal{G}/\mathcal{H}$
then we have
\[
\mathbf{C}_{\;B}^{A}=c_{\;pB}^{A}\boldsymbol{\omega}_{\mathcal{M}}^{p}+c_{\;bB}^{A}\boldsymbol{\omega}_{\mathcal{H}}^{b}
\]
with the condition $c_{\;ab}^{m}=0$ guaranteeing the subgroup condition.
The structure equations for $\tilde{\boldsymbol{\omega}}_{\mathcal{H}}^{a}$
and $\tilde{\boldsymbol{\omega}}_{\mathcal{M}}^{a}$ are 
\begin{eqnarray*}
\mathbf{d}\tilde{\boldsymbol{\omega}}_{\mathcal{H}}^{a} & = & -\frac{1}{2}\mathbf{C}_{\;B}^{a}\wedge\boldsymbol{\omega}^{B}\\
\mathbf{d}\boldsymbol{\tilde{\omega}}_{\mathcal{M}}^{m} & = & -\frac{1}{2}\mathbf{C}_{\;B}^{m}\wedge\boldsymbol{\omega}^{B}
\end{eqnarray*}
Because $c_{\;ab}^{m}=0$, the second of these takes the special form
\begin{eqnarray*}
\mathbf{d}\tilde{\boldsymbol{\omega}}_{\mathcal{M}}^{m} & = & -\frac{1}{2}c_{\;np}^{m}\tilde{\boldsymbol{\omega}}_{\mathcal{M}}^{n}\wedge\tilde{\boldsymbol{\omega}}_{\mathcal{M}}^{p}-c_{\;nb}^{m}\tilde{\boldsymbol{\omega}}_{\mathcal{M}}^{n}\wedge\tilde{\boldsymbol{\omega}}_{\mathcal{H}}^{b}\\
 & = & \left(-\frac{1}{2}c_{\;pn}^{m}\tilde{\boldsymbol{\omega}}_{\mathcal{M}}^{p}-c_{\;bn}^{m}\tilde{\boldsymbol{\omega}}_{\mathcal{H}}^{b}\right)\wedge\tilde{\boldsymbol{\omega}}_{\mathcal{M}}^{n}
\end{eqnarray*}
thereby placing $\tilde{\boldsymbol{\omega}}_{\mathcal{M}}^{m}$ in
involution. This involution means there exist submanifolds found by
setting $\tilde{\boldsymbol{\omega}}_{\mathcal{M}}^{m}=0$, and we
recover the subgroup Lie structure equations of the fibers.

The final step in the construction of a Cartan geometry is to allow
horizontal curvature $2$-forms, $\boldsymbol{\Sigma}^{A}=\left(\boldsymbol{\Sigma}^{a},\boldsymbol{\Omega}^{m}\right)$.
This changes the connection $1$-forms, $\left(\tilde{\boldsymbol{\omega}}_{\mathcal{H}}^{a},\tilde{\boldsymbol{\omega}}_{\mathcal{M}}^{m}\right)$
to a new connection, $\left(\boldsymbol{\omega}_{\mathcal{H}}^{a},\boldsymbol{\omega}_{\mathcal{M}}^{m}\right)$,
and we have the Cartan equations, $\mathbf{d}\boldsymbol{\omega}^{A}=-\frac{1}{2}c_{\;BC}^{A}\boldsymbol{\omega}^{B}\wedge\boldsymbol{\omega}^{C}+\boldsymbol{\Sigma}^{A}$,
or
\begin{eqnarray*}
\mathbf{d}\boldsymbol{\omega}_{\mathcal{H}}^{a} & = & -\frac{1}{2}\mathbf{C}_{\;B}^{a}\wedge\boldsymbol{\omega}^{B}+\boldsymbol{\Sigma}^{a}\\
\mathbf{d}\boldsymbol{\omega}_{\mathcal{M}}^{m} & = & -\frac{1}{2}\mathbf{C}_{\;B}^{m}\wedge\boldsymbol{\omega}^{B}+\boldsymbol{\Omega}^{m}
\end{eqnarray*}
Separating out the subgroup components,
\begin{eqnarray}
\mathbf{d}\boldsymbol{\omega}_{\mathcal{H}}^{a} & = & -\frac{1}{2}c_{\;bc}^{a}\boldsymbol{\omega}_{\mathcal{H}}^{b}\wedge\boldsymbol{\omega}_{\mathcal{H}}^{c}-\frac{1}{2}c_{\;mp}^{a}\boldsymbol{\omega}_{\mathcal{M}}^{m}\wedge\boldsymbol{\omega}_{\mathcal{M}}^{p}-c_{\;mb}^{a}\boldsymbol{\omega}_{\mathcal{M}}^{m}\wedge\boldsymbol{\omega}_{\mathcal{H}}^{b}+\boldsymbol{\Sigma}^{a}\nonumber \\
\mathbf{d}\boldsymbol{\omega}_{\mathcal{M}}^{m} & = & -\frac{1}{2}c_{\;np}^{m}\boldsymbol{\omega}_{\mathcal{M}}^{n}\wedge\boldsymbol{\omega}_{\mathcal{M}}^{p}-c_{\;nb}^{m}\boldsymbol{\omega}_{\mathcal{M}}^{n}\wedge\boldsymbol{\omega}_{\mathcal{H}}^{b}+\boldsymbol{\Omega}^{m}\label{Cartan equations}
\end{eqnarray}
The curvature forms are tensorial under $\mathcal{H}$, and because
they are horizontal, describe curvature of $\mathcal{M}$ only. Consistency
requires the integrability of these equations, which is no longer
guaranteed by the Jacobi identity. Integrability of $\mathbf{d}\boldsymbol{\omega}^{A}=-\frac{1}{2}c_{\;BC}^{A}\boldsymbol{\omega}^{B}\wedge\boldsymbol{\omega}^{C}+\boldsymbol{\Sigma}^{A}$
requires
\begin{eqnarray*}
0 & \equiv & \mathbf{d}^{2}\boldsymbol{\omega}^{A}\\
 & = & -\frac{1}{2}c_{\;BC}^{A}\mathbf{d}\boldsymbol{\omega}^{B}\wedge\boldsymbol{\omega}^{C}+\frac{1}{2}c_{\;BC}^{A}\boldsymbol{\omega}^{B}\wedge\mathbf{d}\boldsymbol{\omega}^{C}+\mathbf{d}\boldsymbol{\Sigma}^{A}\\
 & = & -\frac{1}{2}c_{\;BC}^{A}\left(-\frac{1}{2}c_{\;DE}^{B}\boldsymbol{\omega}^{D}\wedge\boldsymbol{\omega}^{E}+\boldsymbol{\Sigma}^{B}\right)\wedge\boldsymbol{\omega}^{C}+\frac{1}{2}c_{\;BC}^{A}\boldsymbol{\omega}^{B}\wedge\left(-\frac{1}{2}c_{\;DE}^{C}\boldsymbol{\omega}^{D}\wedge\boldsymbol{\omega}^{E}+\boldsymbol{\Sigma}^{C}\right)+\mathbf{d}\boldsymbol{\Sigma}^{A}\\
 & = & \frac{1}{2}c_{\;B[C}^{A}c_{\;DE]}^{B}\boldsymbol{\omega}^{C}\wedge\boldsymbol{\omega}^{D}\wedge\boldsymbol{\omega}^{E}+\mathbf{d}\boldsymbol{\Sigma}^{A}-\frac{1}{2}c_{\;BC}^{A}\boldsymbol{\Sigma}^{B}\wedge\boldsymbol{\omega}^{C}+\frac{1}{2}c_{\;BC}^{A}\boldsymbol{\omega}^{B}\wedge\boldsymbol{\Sigma}^{C}\\
 & = & \mathbf{d}\boldsymbol{\Sigma}^{A}+c_{\;BC}^{A}\boldsymbol{\omega}^{B}\wedge\boldsymbol{\Sigma}^{C}\\
 & \equiv & \mathbf{D}\boldsymbol{\Sigma}^{A}
\end{eqnarray*}
where we use the Jacobi identity, $c_{\;B[C}^{A}c_{\;DE]}^{B}\equiv0$,
and define the covariant exterior derivative. The result holds for
both $\boldsymbol{\Sigma}^{a}$ and $\boldsymbol{\Omega}^{a}$, leaving
us with
\begin{eqnarray*}
\mathbf{D}\boldsymbol{\Sigma}^{a} & = & 0\\
\mathbf{D}\boldsymbol{\Omega}^{a} & = & 0
\end{eqnarray*}
These are the Bianchi identities in gravitational models.

A locally $\mathcal{H}$-invariant physical theory is now found by
writing any scalar Lagrange density built from the tensors available
from the construction of this principal bundle, $\boldsymbol{\mathrm{\Omega}}^{a},\boldsymbol{\Sigma}^{m},$
along with tensors from the original linear representation. The non-vertical
basis forms $\boldsymbol{\omega}_{\mathcal{M}}^{m}$ also become tensors
because the curvature breaks the corresponding symmetries.

\subsection{Conformal structure equations and the Cartan geometry}

Applying the quotient method to the conformal group of a compactified
space with flat metric, $\eta_{ab}$, of signature $\left(p,q\right)$,
we express the generators as 
\begin{eqnarray*}
M_{\;b}^{a} & = & \frac{1}{2}\left(x^{a}\partial_{b}-\eta^{ac}\eta_{bd}x^{d}\partial_{c}\right)\\
 & \equiv & \Delta_{db}^{ac}x^{d}\partial_{c}\\
P_{a} & = & \partial_{a}\\
K^{a} & = & \frac{1}{2}\left(x^{2}\eta^{ab}\partial_{b}-2x^{a}x^{c}\partial_{c}\right)\\
D & = & x^{c}\partial_{c}
\end{eqnarray*}
where the $A$ index on the generators $G_{A}$ includes all possible
antisymmetric pairs for the group, i.e., $A\in\left\{ \left(\begin{array}{c}
a\\
b
\end{array}\right),\left(\begin{array}{c}
\cdot\\
a
\end{array}\right),\left(\begin{array}{c}
a\\
\cdot
\end{array}\right),\left(\begin{array}{c}
\cdot\\
\cdot
\end{array}\right)\right\} $ and $\Delta_{db}^{ac}\equiv\frac{1}{2}\left(\delta_{d}^{a}\delta_{b}^{c}-\eta^{ac}\eta_{bd}\right)$
is the antisymmetric projection operator on $\left(\begin{array}{c}
1\\
1
\end{array}\right)$ tensors. The operators $M_{\;b}^{a},P_{a},K^{a},D$ generate $SO\left(p,q\right)$
transformations, translations, special conformal transformations,
and dilatations, respectively. 

The commutators of the generators then give the Lie algebra, 
\begin{eqnarray*}
\left[M_{\;b}^{a},M_{\;d}^{c}\right] & = & \frac{1}{2}\left[\delta_{b}^{c}\delta_{e}^{a}\delta_{d}^{f}+\eta_{bd}\delta_{e}^{c}\eta^{fa}+\eta^{ac}\eta_{ed}\delta_{b}^{f}+\delta_{d}^{a}\eta_{be}\eta^{fc}\right]M_{\;f}^{e}\\
\left[M_{\;b}^{a},P_{c}\right] & = & \Delta_{cb}^{ad}P_{d}\\
\left[M_{\;b}^{a},K^{c}\right] & = & -\Delta_{db}^{ac}K^{d}\\
\left[P_{a},K^{b}\right] & = & 2\Delta_{ca}^{bd}M_{\;d}^{c}-\delta_{a}^{b}D\\
\left[D,P_{a}\right] & = & -\delta_{a}^{c}P_{c}\\
\left[D,K^{a}\right] & = & \delta_{c}^{a}K^{c}
\end{eqnarray*}
Then we introduce basis $1$-forms dual to the generators
\begin{eqnarray*}
\left\langle M_{\;b}^{a},\tilde{\boldsymbol{\omega}}_{\,\,d}^{c}\right\rangle  & = & 2\Delta_{db}^{ac}\\
\left\langle P_{a},\tilde{\boldsymbol{\omega}}^{b}\right\rangle  & = & \delta_{a}^{b}\\
\left\langle K^{a},\tilde{\boldsymbol{\omega}}_{b}\right\rangle  & = & \delta_{b}^{a}\\
\left\langle D,\tilde{\boldsymbol{\omega}}\right\rangle  & = & 1
\end{eqnarray*}
Notice that index position corresponds to conformal weight, so $\tilde{\boldsymbol{\omega}}^{a}$
and $\tilde{\boldsymbol{\omega}}_{a}$ are distinct fields. These
lead directly to the Maurer-Cartan structure equations 
\begin{eqnarray}
\mathbf{d}\tilde{\boldsymbol{\omega}}_{\,\,b}^{a} & = & \tilde{\boldsymbol{\omega}}_{\,\,b}^{c}\wedge\tilde{\boldsymbol{\omega}}_{\,\,c}^{a}+2\Delta_{cb}^{ad}\tilde{\boldsymbol{\omega}}_{d}\wedge\tilde{\boldsymbol{\omega}}^{c}\nonumber \\
\mathbf{d}\tilde{\boldsymbol{\omega}}^{a} & = & \tilde{\boldsymbol{\omega}}^{c}\wedge\tilde{\boldsymbol{\omega}}_{\,\,c}^{a}+\tilde{\boldsymbol{\omega}}\wedge\tilde{\boldsymbol{\omega}}^{a}\nonumber \\
\mathbf{d}\tilde{\boldsymbol{\omega}}_{a} & = & \tilde{\boldsymbol{\omega}}_{\,\,a}^{c}\wedge\tilde{\boldsymbol{\omega}}_{c}+\tilde{\boldsymbol{\omega}}_{a}\wedge\tilde{\boldsymbol{\omega}}\nonumber \\
\mathbf{d}\tilde{\boldsymbol{\omega}} & = & \tilde{\boldsymbol{\omega}}^{c}\wedge\tilde{\boldsymbol{\omega}}_{c}\label{Maurer-Cartan equations}
\end{eqnarray}
With the quotient $\mathcal{C}/\mathcal{W}$, these equations describe
a $2n$-dimensional homogeneous manifold spanned by $\tilde{\boldsymbol{\omega}}^{c}$
and $\tilde{\boldsymbol{\omega}}_{c}$. Notice that $\mathbf{d}\tilde{\boldsymbol{\omega}}=\tilde{\boldsymbol{\omega}}^{c}\wedge\tilde{\boldsymbol{\omega}}_{c}$
is a symplectic form since it is manifestly both closed and non-degenerate. 

To complete the construction of a Cartan geometry, these connection
forms are now generalized, $\left(\tilde{\boldsymbol{\omega}}_{\,\,b}^{a},\tilde{\boldsymbol{\omega}}^{a},\tilde{\boldsymbol{\omega}}_{a},\tilde{\boldsymbol{\omega}}\right)\rightarrow\left(\boldsymbol{\omega}_{\,\,b}^{a},\boldsymbol{\omega}^{a},\boldsymbol{\omega}_{a},\boldsymbol{\omega}\right)$,
to permit horizontal curvature $2$-forms,
\begin{eqnarray}
\mathbf{d}\boldsymbol{\omega}_{\,\,b}^{a} & = & \boldsymbol{\omega}_{\,\,b}^{c}\wedge\boldsymbol{\omega}_{\,\,c}^{a}+2\Delta_{cb}^{ad}\boldsymbol{\omega}_{d}\wedge\boldsymbol{\omega}^{c}+\boldsymbol{\Omega}_{\;b}^{a}\nonumber \\
\mathbf{d}\boldsymbol{\omega}^{a} & = & \boldsymbol{\omega}^{c}\wedge\boldsymbol{\omega}_{\,\,c}^{a}+\boldsymbol{\omega}\wedge\boldsymbol{\omega}^{a}+\mathbf{T}^{a}\nonumber \\
\mathbf{d}\boldsymbol{\omega}_{a} & = & \boldsymbol{\omega}_{\,\,a}^{c}\wedge\boldsymbol{\omega}_{c}+\boldsymbol{\omega}_{a}\wedge\boldsymbol{\omega}+\mathbf{S}_{a}\nonumber \\
\mathbf{d}\boldsymbol{\omega} & = & \boldsymbol{\omega}^{c}\wedge\boldsymbol{\omega}_{c}+\boldsymbol{\Omega}\label{Cartan equations for C/W}
\end{eqnarray}
The final step in producing a gravity model is to write the most general
action functional linear in the curvatures \cite{WW},
\[
S=\int\left(\alpha\boldsymbol{\Omega}_{\;b}^{a}+\beta\delta_{b}^{a}\boldsymbol{\Omega}+\gamma\boldsymbol{\omega}^{a}\wedge\boldsymbol{\omega}_{b}\right)\varepsilon_{ac\ldots d}^{\qquad be\ldots f}\boldsymbol{\omega}^{d}\wedge\ldots\wedge\boldsymbol{\omega}^{c}\wedge\boldsymbol{\omega}_{e}\wedge\ldots\wedge\boldsymbol{\omega}_{f}
\]
We will be concerned only with the solution to the homogeneous geometry,
eqs.(\ref{Maurer-Cartan equations}), but the complete $\mathbf{T}^{a}=0$
class of solutions has been shown to reduce to locally scale-invariant
general relativity on a Lagrangian submanifold \cite{Torsion free solution}.

For the rest of our discussion, we return to the homogeneous manifold
described by eqs.(\ref{Maurer-Cartan equations}). An extended derivation,
similar to that in \cite{WW,Torsion free solution} but straightforward
to verify by direct substitution, gives the general solution to these
structure equations up to gauge and coordinate choices. The solution
takes the form first given in \cite{NCG}, but a coordinate choice
removes the function $\alpha_{a}\left(x\right)$ present there. Here
we use the symmetry between $\tilde{\boldsymbol{\omega}}^{a}$ and
$\tilde{\boldsymbol{\omega}}_{a}$ to recast that general solution
as 
\begin{eqnarray}
\tilde{\boldsymbol{\omega}}_{\,\,b}^{a} & = & -2\Delta_{db}^{ac}x^{d}\mathbf{d}y_{c}\label{Spin connection}\\
\tilde{\boldsymbol{\omega}}^{a} & = & \mathbf{d}x^{a}-\left(x^{a}x^{b}-\frac{1}{2}\eta^{ab}x^{2}\right)\mathbf{d}y_{b}\label{Solder form}\\
\tilde{\boldsymbol{\omega}}_{a} & = & \mathbf{d}y_{a}\label{Co-solder form}\\
\tilde{\boldsymbol{\omega}} & = & x^{c}\mathbf{d}y_{c}\label{Weyl vector}
\end{eqnarray}
where $x^{2}\equiv\eta_{ab}x^{a}x^{b}$. For convenience, we define
\begin{eqnarray}
b^{ab} & \equiv & 2x^{a}x^{b}-\eta^{ab}x^{2}
\end{eqnarray}
which simplifies eq.(\ref{Co-solder form}) to $\tilde{\boldsymbol{\omega}}^{a}=\mathbf{d}x^{a}-\frac{1}{2}b^{ab}\mathbf{d}y_{b}$.

Our Euclidean starting point gives us a specific model within which
we can explicitly verify and more clearly understand the signature
theorem of \cite{Spencer Wheeler}. Note that when $\eta_{ab}=\delta_{ab}$
is Euclidean, $b^{ab}$ has Lorentzian signature
\[
b^{ab}=-\left|x^{2}\right|\left(\begin{array}{cccc}
-1\\
 & 1\\
 &  & 1\\
 &  &  & 1
\end{array}\right)
\]
but other initial signatures $\left(p,q\right)$ for $\eta_{ab}$
do not give consistent signature for $b_{ab}$. Specifically, for
non-Euclidean $\eta_{ab}$, if $x^{a}$ is timelike $\left(x^{2}<0\right)$,
a transformation to $x^{a}=+\sqrt{\left|x^{2}\right|}\left(1,0,\ldots,0\right)$
gives signature $\left(p',q'\right)=\left(p+1,q-1\right)$ for $b_{ab}$,
while if $x^{a}$ is spacelike $x^{2}>0$, boosting and rotating to
$x^{a}=+\sqrt{\left|x^{2}\right|}\left(0,1,0,0\right)$ at any given
point gives $b_{ab}$ signature $\left(p'',q''\right)=\left(q+1,p-1\right)$.
Equating $\left(p',q'\right)=\left(p'',q''\right)$ requires $p=q$.
Therefore, $b_{ab}$ only has a consistent signature for all $x^{a}$
if $\eta_{ab}$ is Euclidean or if $p=q$ and in the Euclidean case
$b^{ab}$ is necessarily Lorentzian. This is in agreement with the
signature theorem of \cite{Spencer Wheeler}. For the remainder of
this investigation, we will take $\eta_{ab}=\delta_{ab}=diag\left(1,\ldots,1\right)$
so that $b^{ab}=2x^{a}x^{b}-\delta^{ab}x^{2}$ is Lorentzian for all
$x^{a}$.

We show in the next Section that $b_{ab}$ is the restriction of the
Killing metric to certain Lagrangian submanifolds. The argument above
then shows how the uniqueness of the signature theorem occurs. If
the initial signature differs from Euclidean or $p=q$, the restriction
of the Killing metric to the Lagrangian submanifolds is degenerate.

\subsection{The Killing form}

The Killing metric of the biconformal manifold is the restriction
of the conformal Killing form $K_{AB}=tr\left(G_{A}G_{B}\right)$
(equal to $c_{\quad AD}^{C}c_{\quad BC}^{D}$ in the adjoint representation)
to the quotient manifold. The nondegeneracy of $K_{AB}$ allows us
to define the inner product 
\begin{equation}
\left\langle \boldsymbol{\omega}^{A},\boldsymbol{\omega}^{B}\right\rangle \equiv K^{AB}
\end{equation}
in either the curved $\left(\boldsymbol{\omega}^{A}\right)$ or the
homogeneous $\left(\tilde{\boldsymbol{\omega}}^{A}\right)$ case.
Restricting $K^{AB}$ to its form on the quotient manifold, we have
\begin{eqnarray}
\left\langle \boldsymbol{\omega}^{a},\boldsymbol{\omega}^{b}\right\rangle  & = & 0\nonumber \\
\left\langle \boldsymbol{\omega}^{a},\boldsymbol{\omega}_{b}\right\rangle =\left\langle \boldsymbol{\omega}_{b},\boldsymbol{\omega}^{a}\right\rangle  & = & \delta_{b}^{a}\nonumber \\
\left\langle \boldsymbol{\omega}_{a},\boldsymbol{\omega}_{b}\right\rangle  & = & 0\label{Null metric}
\end{eqnarray}
which is readily seen to be nondegenerate. This means that, unlike
the Poincaré case, the group Killing form provides a metric on the
quotient manifold.

It is useful to express this inner product in terms of the $x^{a}$
and $y_{a}$ coordinates as well. Substituting the solution of eqs.(\ref{Spin connection})
- (\ref{Weyl vector}) into the Killing metric, eq.(\ref{Null metric}),
we find the inner product of the coordinate basis forms to be 
\begin{eqnarray*}
\left\langle \mathbf{d}x^{a},\mathbf{d}x^{b}\right\rangle  & = & b^{ab}\\
\left\langle \mathbf{d}x^{a},\mathbf{d}y_{b}\right\rangle =\left\langle \mathbf{d}y_{b},\mathbf{d}x^{a}\right\rangle  & = & \delta_{b}^{a}\\
\left\langle \mathbf{d}y_{a},\mathbf{d}y_{b}\right\rangle  & = & 0
\end{eqnarray*}

In the next Section, we show that there is a Lagrangian submanifold
of flat $SO\left(n\right)$ biconformal space with natural Lorentzian
signature and both $SO\left(n\right)$ and spacetime connections.

\section{An SO(n) Yang-Mills field on spacetime}

We now come to our principal results: the simultaneous presence on
a Lagrangian submanifold of an $SO\left(n\right)$ connection and
its Yang-Mills field strength, together with Lorentzian metric, connection
and curvature. The result is achieved by using a different involution
from that of \cite{Hazboun Wheeler,Hazboun Wheeler-1}.

From the Maurer-Cartan equations, eqs.(\ref{Maurer-Cartan equations}),
we see that both basis forms, $\tilde{\boldsymbol{\omega}}^{a}$ and
separately, $\tilde{\boldsymbol{\omega}}_{a}$, are in involution.
This implies the existence of complementary submanifolds, and the
vanishing of the symplectic form when one or the other basis form
vanishes shows that the submanifolds are Lagrangian. In this Section,
we use a general solution to the flat structure equations to develop
the Killing metric on both the $\tilde{\boldsymbol{\omega}}_{a}=0$
submanifold and the $\tilde{\boldsymbol{\omega}}^{a}=0$ submanifold
and explore properties of the solution, showing the presence of an
$SO\left(n\right)$ Yang-Mills field on spacetime from the elements
of the solution.

\subsection{The structure equations on the spacetime submanifold}

Consider the involution of the basis forms, $\tilde{\boldsymbol{\omega}}^{a}$.
This means that there exist coordinates, found in the next section,
such that $\tilde{\boldsymbol{\omega}}^{a}=\alpha_{\;\;b}^{a}\mathbf{d}u^{b}$,
and holding $u^{a}$ constant sets $\tilde{\boldsymbol{\omega}}^{a}=0$
and selects a submanifold spanned by the remaining basis forms. Setting
$\tilde{\boldsymbol{\omega}}^{a}=0$, eq.(\ref{Solder form}) shows
that $\mathbf{d}y_{a}=2b_{ab}\mathbf{d}x^{b}$ and the structure equations
reduce to 
\begin{eqnarray*}
\tilde{\boldsymbol{\omega}}_{\,\,b}^{a} & = & -\frac{4}{x^{2}}\Delta_{db}^{ac}x_{c}\mathbf{d}x^{d}\\
\tilde{\boldsymbol{\omega}}_{a} & = & 2b_{ab}\mathbf{d}x^{b}\\
\tilde{\boldsymbol{\omega}} & = & \mathbf{d}\left(\ln x^{2}\right)
\end{eqnarray*}
These describe what we will call the spacetime submanifold; it is
a Lagrangian submanifold since the symplectic form $\mathbf{d}\tilde{\boldsymbol{\omega}}=\mathbf{d}^{2}\left(\ln x^{2}\right)\equiv0$
vanishes.

\subsection{Symmetries}

The inner products of the restricted basis forms $\tilde{\boldsymbol{\omega}}_{a}$
are now
\begin{eqnarray*}
\left\langle \tilde{\boldsymbol{\omega}}_{a},\tilde{\boldsymbol{\omega}}_{b}\right\rangle  & = & \left\langle 2b_{ac}\mathbf{d}x^{c},2b_{bd}\mathbf{d}x^{d}\right\rangle \\
 & = & 4b_{ac}b_{bd}\left\langle \mathbf{d}x^{c},\mathbf{d}x^{d}\right\rangle \\
 & = & 4b_{ac}b_{bd}b^{cd}\\
 & = & 4b_{ab}
\end{eqnarray*}
so the basis and coordinate differentials specify a Lorentzian inner
product.

Unlike previous treatments, this metric symmetry is different from
the symmetry of the connection, which is constructed from $SO\left(n\right)$
invariants including the antisymmetric projection $\Delta_{db}^{ac}$.
Since
\begin{eqnarray*}
\Delta_{db}^{ac}\tilde{\boldsymbol{\omega}}_{\,\,c}^{d} & = & -\frac{4}{x^{2}}\Delta_{db}^{ac}\Delta_{fc}^{de}\delta_{eg}x^{g}\mathbf{d}x^{f}\\
 & = & -\frac{4}{x^{2}}\Delta_{fb}^{ae}\delta_{eg}x^{g}\mathbf{d}x^{f}\\
 & = & \tilde{\boldsymbol{\omega}}_{\,\,b}^{a}
\end{eqnarray*}
the infinitesimal change in the Euclidean metric $\delta_{ab}$ produced
by $\tilde{\boldsymbol{\omega}}_{\,\,b}^{a}$ vanishes: 
\[
\delta_{ac}\tilde{\boldsymbol{\omega}}_{\,\,b}^{c}+\delta_{bc}\tilde{\boldsymbol{\omega}}_{\,\,a}^{c}=0
\]
Therefore, $\tilde{\boldsymbol{\omega}}_{\,\,b}^{a}$ is a generator
of $SO\left(n\right)$.

Evidently, though the biconformal bundle as a whole has $SO\left(n\right)$
fiber symmetry, and this symmetry together with dilatations preserve
the structure equations, the symmetry of the metric restricted to
this submanifold is Lorentzian. We develop this further as follows.

Let the $SO\left(n\right)$ gauge be fixed, and consider coordinate
transformations. The vielbein, $\tilde{\boldsymbol{\omega}}_{a}$,
gives rise to a Lorentzian metric:
\begin{eqnarray}
\left\langle \tilde{\boldsymbol{\omega}}_{a},\tilde{\boldsymbol{\omega}}_{b}\right\rangle  & = & 4b_{ab}\label{Lorentzian inner product}
\end{eqnarray}
as do the coordinates
\[
\left\langle \mathbf{d}x^{\alpha},\mathbf{d}x^{\beta}\right\rangle =b^{\alpha\beta}
\]
Given a metric manifold we may construct the frame bundle, $\mathbf{B}\left(\pi,G,\mathcal{M}\right)$.
There then exists the subbundle of orthonormal frames \cite{Isham},
$\mathbf{O}\left(\pi,SO\left(n-1,1\right),\mathcal{M}\right)$, a
principal fiber bundle with Lorentz symmetry group. Therefore, implicit
in the set of general coordinate transformations, we have all Lorentz
transformations of the corresponding orthonormal frame fields. In
this way, we generate a representation of $SO\left(n-1,1\right)$
which is clearly independent of the $SO\left(n\right)$ fiber symmetry.
The bundle structure guarantees that these symmetries are independent,
satisfying the Coleman-Mandula theorem. We regard the combined bundle
as having symmetry $SO\left(n\right)\times SO\left(n-1,1\right)\times SO\left(1,1\right)$.

We define the Christoffel connection, $\Gamma_{\;\mu\nu}^{\alpha}$,
and Riemann curvature, $R_{\;\beta\mu\nu}^{\alpha}$, in the usual
way, from the metric $b_{\alpha\beta}$ and without reference to the
$SO\left(n\right)$ bundle symmetry. The orthogonal bundle symmetry
has its own connection, $\tilde{\boldsymbol{\omega}}_{\,\,b}^{a}$,
and corresponding Yang-Mills field,
\[
\mathbf{F}_{\;b}^{a}=\mathbf{d}\tilde{\boldsymbol{\omega}}_{\,\,b}^{a}-\tilde{\boldsymbol{\omega}}_{\,\,b}^{c}\wedge\tilde{\boldsymbol{\omega}}_{\,\,c}^{a}
\]

This development within the solution of a Lorentzian metric is in
accordance with the signature theorem proved in \cite{Spencer Wheeler}
and developed further in \cite{Hazboun Wheeler,Hazboun Wheeler-1}.

We now compute the spacetime curvature and the Yang-Mills field strength
of this model.

\subsection{Spacetime curvature of the Lagrangian submanifold}

The inverse spacetime metric is
\begin{eqnarray*}
\left\langle \mathbf{d}x^{\alpha},\mathbf{d}x^{\beta}\right\rangle  & = & b^{\alpha\beta}\\
 & = & 2x^{\alpha}x^{\beta}-\delta^{\alpha\beta}x^{2}
\end{eqnarray*}
In these coordinates, $\delta^{\alpha\beta}=diag\left(1,\ldots,1\right)$.
Inverting,
\begin{eqnarray*}
g_{\alpha\beta} & = & \frac{1}{\left(x^{2}\right)^{2}}\left(2x_{\alpha}x_{\beta}-\delta_{\alpha\beta}x^{2}\right)
\end{eqnarray*}
The Christoffel connection is readily found to be,
\begin{eqnarray*}
\Gamma_{\;\beta\mu}^{\alpha} & = & \frac{1}{x^{2}}\left(x^{\alpha}\delta_{\beta\mu}-\delta_{\beta}^{\alpha}x_{\mu}-\delta_{\mu}^{\alpha}x_{\beta}\right)
\end{eqnarray*}
and the curvature is
\begin{eqnarray*}
R_{\;\beta\mu\nu}^{\alpha} & = & \Gamma_{\;\beta\nu,\mu}^{\alpha}-\Gamma_{\;\beta\mu,\nu}^{\alpha}-\Gamma_{\;\sigma\nu}^{\alpha}\Gamma_{\;\beta\mu}^{\sigma}+\Gamma_{\;\sigma\mu}^{\alpha}\Gamma_{\;\beta\nu}^{\sigma}\\
 & = & \frac{1}{x^{2}}\left(\left(\delta_{\mu}^{\alpha}-\frac{1}{x^{2}}x^{\alpha}x_{\mu}\right)\left(\delta_{\beta\nu}-\frac{1}{x^{2}}x_{\beta}x_{\nu}\right)-\left(\delta_{\nu}^{\alpha}-\frac{1}{x^{2}}x^{\alpha}x_{\nu}\right)\left(\delta_{\beta\mu}-\frac{1}{x^{2}}x_{\beta}x_{\mu}\right)\right)
\end{eqnarray*}
Defining the projection operator $P_{\;\;\mu}^{\alpha}\equiv\left(\delta_{\mu}^{\alpha}-\frac{1}{x^{2}}x^{\alpha}x_{\mu}\right)$,
orthogonal to $x^{\alpha}$, we may write this as
\begin{eqnarray*}
R_{\;\beta\mu\nu}^{\alpha} & = & \frac{1}{x^{2}}\left(P_{\;\;\mu}^{\alpha}P_{\beta\nu}-P_{\;\;\nu}^{\alpha}P_{\beta\mu}\right)
\end{eqnarray*}

\subsection{The Yang-Mills field strength}

Finally, we compute the Yang-Mills field strength,
\[
\mathbf{F}_{\;b}^{a}=\mathbf{d}\tilde{\boldsymbol{\omega}}_{\,\,b}^{a}-\tilde{\boldsymbol{\omega}}_{\,\,b}^{c}\wedge\tilde{\boldsymbol{\omega}}_{\,\,c}^{a}
\]
where the $SO\left(n\right)$ connection is given by
\[
\tilde{\boldsymbol{\omega}}_{\,\,b}^{a}=-\frac{4}{x^{2}}\Delta_{db}^{ac}x_{c}\mathbf{d}x^{d}
\]
We may find this directly from the $SO\left(n\right)$ structure equation,
\begin{eqnarray*}
\mathbf{d}\tilde{\boldsymbol{\omega}}_{\,\,b}^{a} & = & \tilde{\boldsymbol{\omega}}_{\,\,b}^{c}\wedge\tilde{\boldsymbol{\omega}}_{\,\,c}^{a}+2\Delta_{cb}^{ad}\tilde{\boldsymbol{\omega}}_{d}\wedge\tilde{\boldsymbol{\omega}}^{c}
\end{eqnarray*}
which shows immediately that
\[
\mathbf{F}_{\;b}^{a}=-\left.2\Delta_{cb}^{ad}\tilde{\boldsymbol{\omega}}_{d}\wedge\tilde{\boldsymbol{\omega}}^{c}\right|_{\tilde{\boldsymbol{\omega}}^{c}=0}=0
\]
The Yang-Mills field strength therefore vanishes on the spacetime
Lagrangian submanifold.

We easily check this directly. Substituting the submanifold form,
$-\frac{4}{x^{2}}\Delta_{db}^{ac}x_{c}\mathbf{d}x^{d}$ leads after
some algebra to $\mathbf{F}_{\;b}^{a}=0$ (see Appendix 1). Naturally,
this will change when curved biconformal spaces are considered.

We end the section by finding a set of coordinates adapted to the
involution of $\tilde{\boldsymbol{\omega}}^{a}$.

\subsection{Adapted coordinates for the involution of the translational gauge
fields}

The involution of $\boldsymbol{\omega}^{a}$ means that we may write
$\boldsymbol{\omega}^{a}=\alpha_{\;\;\beta}^{\alpha}\mathbf{d}u^{\beta}$
for $n$ coordinates $u^{\alpha}$. To find $\alpha_{\;\;\beta}^{\alpha}$
and $u^{\beta}$, we need to solve
\begin{equation}
\alpha_{\;\;\beta}^{\alpha}\mathbf{d}u^{\beta}=\mathbf{d}x^{\alpha}-\frac{1}{2}b^{\alpha\beta}\mathbf{d}y_{\beta}\label{Involution coordinates}
\end{equation}
With the metric $b_{\alpha\beta}$ given by
\begin{eqnarray*}
b_{\alpha\beta} & = & \frac{1}{\left(x^{2}\right)^{2}}\left(2x_{\alpha}x_{\beta}-\delta_{\alpha\beta}x^{2}\right)
\end{eqnarray*}
where we define $x_{\alpha}\equiv\delta_{\alpha\beta}x^{\beta}$,
we rewrite eq.(\ref{Involution coordinates}) as $b_{\alpha\mu}\alpha_{\;\;\beta}^{\mu}\mathbf{d}u^{\beta}=b_{\alpha\mu}\mathbf{d}x^{\mu}-\frac{1}{2}\mathbf{d}y_{\alpha}$
and note that $b_{\alpha\mu}\mathbf{d}x^{\mu}=-\delta_{\alpha\mu}\mathbf{d}\left(\frac{x^{\mu}}{x^{2}}\right)$.
Therefore,
\begin{eqnarray*}
2b_{\alpha\mu}\alpha_{\;\;\beta}^{\mu}\mathbf{d}u^{\beta} & = & -\delta_{\alpha\mu}\mathbf{d}\left(\frac{2x^{\mu}}{x^{2}}+\delta^{\mu\beta}y_{\beta}\right)
\end{eqnarray*}
and we may identify
\begin{eqnarray*}
u^{\alpha} & \equiv & \frac{2x^{\alpha}}{x^{2}}+\delta^{\alpha\beta}y_{\beta}
\end{eqnarray*}
and require
\begin{eqnarray*}
2b_{\alpha\mu}\alpha_{\;\;\beta}^{\mu} & = & \delta_{\alpha\beta}
\end{eqnarray*}
Inverting the metric on this expression now shows that
\[
\alpha_{\;\;\beta}^{\mu}=x^{\alpha}x_{\beta}-\frac{1}{2}\delta_{\beta}^{\alpha}x^{2}
\]

In terms of the adapted coordinates $u^{\alpha}$, we may solve for
and replace $x^{\alpha}$,
\begin{eqnarray*}
x^{\alpha}\left(u^{\alpha},y_{\beta}\right) & = & \frac{2\left(u^{\alpha}-\delta^{\alpha\beta}y_{\beta}\right)}{u^{2}-2u^{\alpha}y_{\alpha}+y^{2}}
\end{eqnarray*}
so that the full structure equations take the form
\begin{eqnarray}
\tilde{\boldsymbol{\omega}}_{\,\,b}^{a} & = & -2\Delta_{db}^{ac}\delta_{\mu}^{d}\delta_{c}^{\nu}x^{\mu}\left(u^{\alpha},y_{\beta}\right)\mathbf{d}y_{\nu}\nonumber \\
\tilde{\boldsymbol{\omega}}^{a} & = & \delta_{\beta}^{a}\mathbf{d}u^{\beta}\nonumber \\
\tilde{\boldsymbol{\omega}}_{a} & = & \delta_{a}^{\beta}\mathbf{d}y_{\beta}\nonumber \\
\tilde{\boldsymbol{\omega}} & = & x^{\mu}\left(u^{\alpha},y_{\beta}\right)\mathbf{d}y_{\mu}\label{Structure equations in adapted basis}
\end{eqnarray}
and we note that the second involution, for $\tilde{\boldsymbol{\omega}}_{a}$,
is simultaneously expressed in adapted coordinates $y_{\alpha}$.
It is straightforward to check both directly and in terms of the $\left(x^{\alpha},y_{\beta}\right)$
inner products that $\left\langle \mathbf{d}u^{\alpha},\mathbf{d}u^{\beta}\right\rangle =0$.

\section{Phase space}

The full biconformal space is a symplectic manifold, but unlike our
usual basis for phase spaces, the inner product of th $\left(\tilde{\boldsymbol{\omega}}^{a},\tilde{\boldsymbol{\omega}}_{a}\right)$
basis is off-diagonal, eq.(\ref{Null metric}). To make a full connection
between the background biconformal manifold and a one-particle phase
space, we introduce a further change of basis.

\subsection{An orthogonal basis of Lagrangian submanifolds}

Having both metric and symplectic form, we may find a subspace orthogonal
to the spacetime manifold. To find it, we introduce new coordinates
$z_{a}$ such that $\left\langle \mathbf{d}z_{a},\mathbf{d}x^{b}\right\rangle =0$.
Making the ansatz $\mathbf{d}z_{a}=\xi_{a}^{\quad b}\mathbf{d}y_{b}+\Xi_{ab}\mathbf{d}x^{b}$,
we solve for $\xi_{a}^{\quad b}$ and $\Xi_{ab}$ 
\begin{eqnarray*}
0 & = & \left\langle \mathbf{d}z_{a},\mathbf{d}x^{c}\right\rangle \\
 & = & \xi_{a}^{\quad b}\delta_{b}^{c}+\Xi_{ab}b^{bc}
\end{eqnarray*}
This is satisfied if $\xi_{a}^{c}=\delta_{a}^{c}$ and $\Xi_{ab}=-b_{ab},$
giving
\[
\mathbf{d}z_{a}=\mathbf{d}y_{a}-b_{ab}\mathbf{d}x^{b}
\]
for the differential of the coordinates.

These subspaces are better aligned with our usual notion of a single
particle phase space. In addition to the inner product of the spacetime
coordinate differentials $\left\langle \mathbf{d}x^{a},\mathbf{d}x^{b}\right\rangle =b^{ab}$,
and orthogonality of the spacetime and momentum directions, we have
a metric on the momentum space as well, 
\begin{eqnarray*}
\left\langle \mathbf{d}z_{a},\mathbf{d}z_{b}\right\rangle  & = & \left\langle \mathbf{d}y_{a}-b_{ac}\mathbf{d}x^{c},\mathbf{d}y_{b}-b_{bd}\mathbf{d}x^{d}\right\rangle \\
 & = & -b_{ab}
\end{eqnarray*}
In terms of the $\left(x^{\alpha},z_{\beta}\right)$ coordinates,
the solution for the connection is
\begin{eqnarray*}
\tilde{\boldsymbol{\omega}}_{\,\,b}^{a} & = & -2\Delta_{db}^{ac}x^{d}\left(\mathbf{d}z_{c}+b_{ce}\mathbf{d}x^{e}\right)\\
\tilde{\boldsymbol{\omega}}^{a} & = & \frac{1}{2}\left(\mathbf{d}x^{a}-b^{ab}\mathbf{d}z_{b}\right)\\
\tilde{\boldsymbol{\omega}}_{a} & = & \mathbf{d}z_{a}+b_{ac}\mathbf{d}x^{c}\\
\tilde{\boldsymbol{\omega}} & = & x^{a}\mathbf{d}z_{a}+\frac{1}{x^{2}}x_{a}\mathbf{d}x^{a}
\end{eqnarray*}
However, in order for $\left(x^{\alpha},z_{\beta}\right)$ coordinates
to fully match our expectations for canonically conjugate variables,
spacetime and momentum space must be Lagrangian submanifolds. Therefore
the differentials $\mathbf{d}x^{\alpha}$ and $\mathbf{d}z_{\alpha}$
need to be integrable. In general, this is not the case. If we substitute
$\mathbf{d}x^{a}\rightarrow\boldsymbol{\kappa}^{a}$ and $\mathbf{d}z_{a}\rightarrow\boldsymbol{\lambda}^{a}$
in the structure equations and solve for $\mathbf{d}\boldsymbol{\kappa}^{a}$
and $\mathbf{d}\boldsymbol{\lambda}^{a}$, they are not involute.
For $\mathbf{d}\boldsymbol{\kappa}^{a}$ , for example, we find
\begin{eqnarray*}
2\mathbf{d}\boldsymbol{\kappa}^{a} & = & \boldsymbol{\kappa}^{c}\wedge\tilde{\boldsymbol{\omega}}_{\,\,c}^{a}+\boldsymbol{\kappa}^{a}\wedge\tilde{\boldsymbol{\omega}}-b^{ac}\mathbf{d}b_{cb}\boldsymbol{\kappa}^{b}+\tilde{\boldsymbol{\omega}}\wedge\boldsymbol{\kappa}^{a}+b^{ae}\tilde{\boldsymbol{\omega}}_{\,\,e}^{c}\wedge b_{cb}\boldsymbol{\kappa}^{b}\\
 &  & -b^{cb}\boldsymbol{\lambda}_{b}\wedge\tilde{\boldsymbol{\omega}}_{\,\,c}^{a}-\tilde{\boldsymbol{\omega}}\wedge b^{ab}\boldsymbol{\lambda}_{b}+\mathbf{d}b^{ab}\boldsymbol{\lambda}_{b}+b^{ab}\tilde{\boldsymbol{\omega}}_{\,\,b}^{c}\wedge\boldsymbol{\lambda}_{c}+b^{ab}\boldsymbol{\lambda}_{b}\wedge\tilde{\boldsymbol{\omega}}
\end{eqnarray*}
Therefore, in order for $\mathbf{d}x^{a}$ and $\mathbf{d}z_{a}$
to span a pair of Lagrangian submanifolds, we require the $\boldsymbol{\lambda}_{a}\wedge\boldsymbol{\lambda}_{b}$
terms to vanish. Equivalently, we must have
\[
\boldsymbol{\Lambda}^{a}\equiv-b^{cb}\boldsymbol{\lambda}_{b}\wedge\tilde{\boldsymbol{\omega}}_{\,\,c}^{a}-\tilde{\boldsymbol{\omega}}\wedge b^{ab}\boldsymbol{\lambda}_{b}+\mathbf{d}b^{ab}\boldsymbol{\lambda}_{b}+b^{ab}\tilde{\boldsymbol{\omega}}_{\,\,b}^{c}\wedge\boldsymbol{\lambda}_{c}+b^{ab}\boldsymbol{\lambda}_{b}\wedge\tilde{\boldsymbol{\omega}}
\]
linear in $\boldsymbol{\kappa}^{a}$, with a similar condition $\boldsymbol{\Sigma}^{a}\sim\boldsymbol{\lambda}^{a}$arising
for the involution of $\mathbf{d}\boldsymbol{\lambda}_{b}$.

As we show in detail in Appendix 2, these conditions \emph{do} hold
for the homogeneous solution above – substituting the form of the
connection into $\boldsymbol{\Lambda}^{a}$ and $\boldsymbol{\Sigma}^{a}$
so in the model considered $\left(x^{\alpha},z_{\beta}\right)$ do
characterize Lagrangian submanifolds.

Restricting to the constant $z_{\alpha}$ submanifold, the solution
is
\begin{eqnarray*}
\tilde{\boldsymbol{\omega}}_{\,\,b}^{a} & = & -2\Delta_{db}^{ac}x^{d}b_{ce}\mathbf{d}x^{e}\\
\tilde{\boldsymbol{\omega}}^{a} & = & \frac{1}{2}\mathbf{d}x^{a}\\
\tilde{\boldsymbol{\omega}} & = & \frac{1}{x^{2}}x_{a}\mathbf{d}x^{a}=\frac{1}{2}\mathbf{d}\ln x^{2}
\end{eqnarray*}
with metric $4b_{ab}$ and $SO\left(4\right)$ connection and $\tilde{\boldsymbol{\omega}}_{a}=2b_{ac}\tilde{\boldsymbol{\omega}}^{c}$,
as before. If instead, we hold $x^{\alpha}$ constant, then the solution
is
\begin{eqnarray*}
\tilde{\boldsymbol{\omega}}_{\,\,b}^{a} & = & -2\Delta_{db}^{ac}x_{0}^{d}\mathbf{d}z_{c}\\
\tilde{\boldsymbol{\omega}}_{a} & = & \mathbf{d}z_{a}\\
\tilde{\boldsymbol{\omega}} & = & x_{0}^{a}\mathbf{d}z_{a}
\end{eqnarray*}
with $\tilde{\boldsymbol{\omega}}^{a}=-\frac{1}{2}b^{ab}\mathbf{d}z_{b}$,
\emph{constant} metric $-b^{ab}\left(x_{0}^{c}\right)$ and constant
$SO\left(4\right)$ connection.

While the constancy of the $SO\left(4\right)$ connection does not
imply vanishing of the Yang-Mills field, it may be gauged to zero
by a conformal transformation. The constancy of the metric, $SO\left(4\right)$
connection, and Weyl vector show the momentum space to be a Lagrangian
submanifold with vanishing curvature and vanishing Yang-Mills field.
It is therefore consistent to identify the entire biconformal manifold
with the co-tangent bundle.

It is unclear which of these properties survive in curved biconformal
spaces. Torsion-free biconformal spaces are known to be fully determined
by the spacetime solder form and Weyl vector with $\boldsymbol{\omega}_{a}$
spanning the co-tangent spaces, but the solutions apply to different
submanifolds than we consider in this Section. Still, it is conceivable
that the co-tangent interpretation is always possible.

It is amusing to speculate on the meaning of the momentum space if
it is found to be curved in some solutions. If so, it might provide
a novel approach to canonical field theory by allowing canonically
conjugate fields to co-exist on what is essentially a particle phase
space. Expressing a relativistic field theory on a particle phase
space has had only measured success. Born \cite{Born} suggested introducing
a curved momentum space to complement gravitating spacetime, but with
no clear indication of what would determine its curvature. In the
non-relativistic case, the Wigner distribution extends the wave function
to a distribution on phase space, but it is not obvious how this generalizes
to the relativistic case. Typically, phase space for field theory
employs a natural symplectic structure on field space, but what occurs
here seems to exist midway between the particle and field cases. The
symplectic base manifold allows fields to acquire a momentum component
automatically, and the field equations determine the structure of
the entire phase space, apparently restricting these fields so that
the only independent degrees of freedom are those from the spacetime
Lagrangian submanifold.

\section{Conclusion}

The Coleman-Mandula theorem shows that unifying gauge theories that
include gravity and are based on Lie groups require a direct product
between the internal and gravitational symmetries. This seemed to
be an unnatural starting point for a unified theory, and led to an
increased emphasis on supersymmetric theories. These avoid being direct
products by allowing graded Lie groups, which in turn give improved
quantum convergence and useful restrictions on possible models.

In the present work, we show that it is possible to write a unified
theory as a gauge theory of a simple Lie group which dynamically enforces
the Coleman-Mandula theorem. Although the starting point is the conformal
group $SO\left(n-1,1\right)$ of Euclidean $n$-space, the general
solution of the Maurer-Cartan structure equations shows that the connection
retains its original $SO\left(n\right)$ symmetry but the Killing
metric (which is non-degenerate in these models) restricts to a Lorentzian
signature on certain Lagrangian submanifolds. 

We outlined the quotient manifold method and applied it to the biconformal
gauging of the conformal group. By starting from the generators, we
constructed the Maurer-Cartan structure equations for the conformal
group. The known solution to these equations was introduced, though
we choose to use different coordinates better adapted to the Lagrangian
submanifolds.

Our principal contribution was to identify a complementary pair of
Lagrangian submanifolds on which the connection (of the principal
fiber bundle) is orthogonal ($SO\left(n\right)$) while the restriction
of the Killing metric is Lorentzian. The Lorentzian metric allows
calculation of its Christoffel connection and curvature, while the
$SO\left(n\right)$ connection gives rise to a Yang-Mills field (trivial,
since we only consider the flat case). The Riemannian curvature of
the submanifolds was found to be constructible from projection operators
orthogonal to the time direction. Notice that the time direction emerges
dynamically in accordance with the signature theorem of \cite{Spencer Wheeler}
and further developed in \cite{Hazboun Wheeler}.

By the definition of a fiber bundle, the $SO\left(n\right)$ fibers
are in a direct product with the $SO(n-1,1)$ symmetry of the base
manifold. This direct product relation naturally satisfies the Coleman-Mandula
theorem.

We showed that the full biconformal space may be given the structure
of a co-tangent bundle to the curved spacetime.

Preliminary work with curved biconformal spaces suggests that these
results extend to those cases as well, though the computations pose
some interesting challenges. Since these constructions depend only
on the Lie algebra they will work in the same way with spinor representations.
There is no obstruction to considering supersymmetric generalizations
\cite{AW}, which among other benefits provide a principled way of
introducing spinor fields into bosonic field theories.

\section*{Appendix 1: Vanishing of the submanifold Yang-Mills field}

The $SO\left(n\right)$ Yang-Mills field strength is
\[
\mathbf{F}_{\;b}^{a}=-\Delta_{cb}^{ad}b^{cf}\left(\mathbf{d}z_{d}\wedge\mathbf{d}z_{f}+2b_{de}\mathbf{d}x^{e}\wedge\mathbf{d}z_{f}\right)
\]
which vanishes on the $z_{a}=constant$ Lagrangian submanifold. To
verify this explicitly, we substitute the submanifold form of the
connection, $\left.\tilde{\boldsymbol{\omega}}_{\,\,b}^{a}\right|_{z_{a}=z_{a}^{0}}=-\frac{4}{x^{2}}\Delta_{db}^{ac}x_{c}\mathbf{d}x^{d}$
into $\mathbf{F}_{\;b}^{a}=\mathbf{d}\tilde{\boldsymbol{\omega}}_{\,\,b}^{a}-\tilde{\boldsymbol{\omega}}_{\,\,b}^{c}\wedge\tilde{\boldsymbol{\omega}}_{\,\,c}^{a}$:

\begin{eqnarray*}
\mathbf{F}_{\;b}^{a} & = & \mathbf{d}\left(-\frac{4}{x^{2}}\Delta_{db}^{ac}x_{c}\mathbf{d}x^{d}\right)-\frac{4}{x^{2}}\Delta_{db}^{ce}x_{e}\mathbf{d}x^{d}\wedge\frac{4}{x^{2}}\Delta_{gc}^{af}x_{f}\mathbf{d}x^{g}\\
 & = & \frac{8}{\left(x^{2}\right)^{2}}\Delta_{db}^{ac}x_{e}x_{c}\mathbf{d}x^{e}\wedge\mathbf{d}x^{d}-\frac{4}{x^{2}}\Delta_{db}^{ac}\mathbf{d}x_{c}\wedge\mathbf{d}x^{d}-\frac{16}{\left(x^{2}\right)^{2}}\Delta_{db}^{ce}\Delta_{gc}^{af}x_{e}x_{f}\mathbf{d}x^{d}\wedge\mathbf{d}x^{g}\\
 & = & \frac{8}{\left(x^{2}\right)^{2}}\Delta_{db}^{ac}x_{e}x_{c}\mathbf{d}x^{e}\wedge\mathbf{d}x^{d}-\frac{4}{x^{2}}\Delta_{db}^{ac}\delta_{ce}\mathbf{d}x^{e}\wedge\mathbf{d}x^{d}-\frac{4}{\left(x^{2}\right)^{2}}\left(\delta_{d}^{c}\delta_{b}^{e}-\delta^{ce}\delta_{bd}\right)\left(\delta_{g}^{a}\delta_{c}^{f}-\delta^{af}\delta_{gc}\right)x_{e}x_{f}\mathbf{d}x^{d}\wedge\mathbf{d}x^{g}\\
 & = & \frac{8}{\left(x^{2}\right)^{2}}\Delta_{db}^{ac}\left(x_{e}x_{c}-\frac{1}{2}x^{2}\delta_{ce}\right)\mathbf{d}x^{e}\wedge\mathbf{d}x^{d}\\
 &  & -\frac{4}{\left(x^{2}\right)^{2}}\left(\delta_{e}^{a}x_{b}x_{d}+\delta_{bd}x_{e}x^{a}-x^{2}\delta_{bd}\delta_{e}^{a}\right)\mathbf{d}x^{d}\wedge\mathbf{d}x^{e}\\
 & = & \frac{8}{\left(x^{2}\right)^{2}}\Delta_{db}^{ac}\left(x_{e}x_{c}-\frac{1}{2}x^{2}\delta_{ce}\right)\mathbf{d}x^{e}\wedge\mathbf{d}x^{d}\\
 &  & -\frac{4}{\left(x^{2}\right)^{2}}\left(\left(x_{b}x_{d}-\frac{1}{2}x^{2}\delta_{bd}\right)\delta_{e}^{a}+\left(x^{a}x_{e}-\frac{1}{2}x^{2}\delta_{e}^{a}\right)\delta_{bd}\right)\mathbf{d}x^{d}\wedge\mathbf{d}x^{e}\\
 & = & \frac{8}{\left(x^{2}\right)^{2}}\Delta_{db}^{ac}\left(x_{e}x_{c}-\frac{1}{2}x^{2}\delta_{ce}\right)\mathbf{d}x^{e}\wedge\mathbf{d}x^{d}\\
 &  & -\frac{4}{\left(x^{2}\right)^{2}}\left(x_{c}x_{d}-\frac{1}{2}x^{2}\delta_{cd}\right)\left(\delta_{b}^{c}\delta_{e}^{a}-\delta^{ac}\delta_{be}\right)\mathbf{d}x^{d}\wedge\mathbf{d}x^{e}\\
 & = & \frac{4}{\left(x^{2}\right)^{2}}\Delta_{db}^{ac}\left(2x_{e}x_{c}-x^{2}\delta_{ce}\right)\mathbf{d}x^{e}\wedge\mathbf{d}x^{d}-\frac{4}{\left(x^{2}\right)^{2}}\Delta_{eb}^{ac}\left(2x_{c}x_{d}-x^{2}\delta_{cd}\right)\mathbf{d}x^{d}\wedge\mathbf{d}x^{e}\\
 & = & \frac{4}{\left(x^{2}\right)^{2}}\Delta_{db}^{ac}\left(2x_{e}x_{c}-x^{2}\delta_{ce}\right)\mathbf{d}x^{e}\wedge\mathbf{d}x^{d}-\frac{4}{\left(x^{2}\right)^{2}}\Delta_{db}^{ac}\left(2x_{c}x_{e}-x^{2}\delta_{cd}\right)\mathbf{d}x^{e}\wedge\mathbf{d}x^{d}\\
 & = & 0
\end{eqnarray*}

\section*{Appendix 2: Involution of the orthogonal subspaces}

We find the involution conditions for the $\left(\mathbf{d}x^{a},\mathbf{d}u_{b}\right)$
basis and show that they are satisfied by the homogeneous solution. 

The $\left(\mathbf{d}x^{a},\mathbf{d}u_{b}\right)$ basis is related
to the original basis by
\begin{eqnarray*}
\tilde{\boldsymbol{\omega}}^{a} & = & \frac{1}{2}\left(\mathbf{d}x^{a}-b^{ab}\mathbf{d}u_{b}\right)\\
\tilde{\boldsymbol{\omega}}_{a} & = & \mathbf{d}u_{a}+b_{ab}\mathbf{d}x^{b}
\end{eqnarray*}
If we write this instead in the form of an alternative basis,
\begin{eqnarray*}
\tilde{\boldsymbol{\omega}}^{a} & = & \frac{1}{2}\left(\boldsymbol{\kappa}^{a}-b^{ab}\boldsymbol{\lambda}_{b}\right)\\
\tilde{\boldsymbol{\omega}}_{a} & = & \boldsymbol{\lambda}_{a}+b_{ab}\boldsymbol{\kappa}^{b}
\end{eqnarray*}
and substitute into the original structure equations,
\begin{eqnarray*}
\mathbf{d}\tilde{\boldsymbol{\omega}}^{a} & = & \tilde{\boldsymbol{\omega}}^{c}\wedge\tilde{\boldsymbol{\omega}}_{\,\,c}^{a}+\tilde{\boldsymbol{\omega}}\wedge\tilde{\boldsymbol{\omega}}^{a}\\
\mathbf{d}\tilde{\boldsymbol{\omega}}_{a} & = & \tilde{\boldsymbol{\omega}}_{\,\,a}^{c}\wedge\tilde{\boldsymbol{\omega}}_{c}+\tilde{\boldsymbol{\omega}}_{a}\wedge\tilde{\boldsymbol{\omega}}
\end{eqnarray*}
we find the structure equations for $\boldsymbol{\kappa}^{a}$ and
$\boldsymbol{\lambda}^{a}$,
\begin{eqnarray*}
\mathbf{d}\boldsymbol{\kappa}^{a} & = & \frac{1}{2}\left(\boldsymbol{\kappa}^{c}\wedge\tilde{\boldsymbol{\omega}}_{\,\,c}^{a}+\boldsymbol{\kappa}^{a}\wedge\tilde{\boldsymbol{\omega}}-b^{ac}\mathbf{d}b_{cb}\boldsymbol{\kappa}^{b}+\tilde{\boldsymbol{\omega}}\wedge\boldsymbol{\kappa}^{a}+b^{ae}\tilde{\boldsymbol{\omega}}_{\,\,e}^{c}\wedge b_{cb}\boldsymbol{\kappa}^{b}\right)\\
 &  & +\frac{1}{2}\left(-b^{cb}\boldsymbol{\lambda}_{b}\wedge\tilde{\boldsymbol{\omega}}_{\,\,c}^{a}-\tilde{\boldsymbol{\omega}}\wedge b^{ab}\boldsymbol{\lambda}_{b}+\mathbf{d}b^{ab}\boldsymbol{\lambda}_{b}+b^{ab}\tilde{\boldsymbol{\omega}}_{\,\,b}^{c}\wedge\boldsymbol{\lambda}_{c}+b^{ab}\boldsymbol{\lambda}_{b}\wedge\tilde{\boldsymbol{\omega}}\right)\\
\mathbf{d}\boldsymbol{\lambda}_{a} & = & \frac{1}{2}\left(b_{ad}b^{cb}\boldsymbol{\lambda}_{b}\wedge\tilde{\boldsymbol{\omega}}_{\,\,c}^{d}+\tilde{\boldsymbol{\omega}}\wedge\boldsymbol{\lambda}_{a}-b_{ac}\mathbf{d}b^{cb}\boldsymbol{\lambda}_{b}+\tilde{\boldsymbol{\omega}}_{\,\,a}^{c}\wedge\boldsymbol{\lambda}_{c}+\boldsymbol{\lambda}_{a}\wedge\tilde{\boldsymbol{\omega}}\right)\\
 &  & +\frac{1}{2}\left(\tilde{\boldsymbol{\omega}}_{\,\,a}^{c}\wedge b_{cb}\boldsymbol{\kappa}^{b}+b_{ac}\boldsymbol{\kappa}^{c}\wedge\tilde{\boldsymbol{\omega}}-\mathbf{d}b_{ab}\boldsymbol{\kappa}^{b}-b_{ab}\boldsymbol{\kappa}^{c}\wedge\tilde{\boldsymbol{\omega}}_{\,\,c}^{b}-b_{ab}\tilde{\boldsymbol{\omega}}\wedge\boldsymbol{\kappa}^{b}\right)
\end{eqnarray*}
Therefore, involution of the new basis requires
\[
\boldsymbol{\Lambda}^{a}\equiv-b^{cb}\boldsymbol{\lambda}_{b}\wedge\tilde{\boldsymbol{\omega}}_{\,\,c}^{a}-\tilde{\boldsymbol{\omega}}\wedge b^{ab}\boldsymbol{\lambda}_{b}+\mathbf{d}b^{ab}\boldsymbol{\lambda}_{b}+b^{ab}\tilde{\boldsymbol{\omega}}_{\,\,b}^{c}\wedge\boldsymbol{\lambda}_{c}+b^{ab}\boldsymbol{\lambda}_{b}\wedge\tilde{\boldsymbol{\omega}}
\]
to be proportional to $\boldsymbol{\kappa}^{a}$ and
\[
\boldsymbol{\Sigma}^{a}\equiv b^{ae}\tilde{\boldsymbol{\omega}}_{\,\,e}^{c}\wedge b_{cb}\boldsymbol{\kappa}^{b}+\boldsymbol{\kappa}^{a}\wedge\tilde{\boldsymbol{\omega}}-b^{ac}\mathbf{d}b_{cb}\boldsymbol{\kappa}^{b}-\boldsymbol{\kappa}^{c}\wedge\tilde{\boldsymbol{\omega}}_{\,\,c}^{a}-\tilde{\boldsymbol{\omega}}\wedge\boldsymbol{\kappa}^{a}
\]
to be proportional to $\boldsymbol{\lambda}^{a}$.

The involution does not generically hold, but it only needs to hold
for the solution. Expressing the solution in terms of $\left(\boldsymbol{\kappa}^{a},\boldsymbol{\lambda}_{b}\right)$,
\begin{eqnarray*}
\tilde{\boldsymbol{\omega}}_{\,\,b}^{a} & = & -2\Delta_{db}^{ac}\left(x^{d}\mathbf{d}u_{c}-b_{cf}x^{d}\mathbf{d}x^{f}+\frac{2}{x^{2}}x_{c}\mathbf{d}x^{d}\right)\\
 & = & -2\Delta_{db}^{ac}\left(x^{d}\boldsymbol{\lambda}_{c}-b_{cf}x^{d}\boldsymbol{\kappa}^{f}+\frac{2}{x^{2}}x_{c}\boldsymbol{\kappa}^{d}\right)\\
\tilde{\boldsymbol{\omega}}^{a} & = & \frac{1}{2}\left(\boldsymbol{\kappa}^{a}-b^{ab}\boldsymbol{\lambda}_{b}\right)\\
\tilde{\boldsymbol{\omega}}_{a} & = & \boldsymbol{\lambda}_{a}+b_{ab}\boldsymbol{\kappa}^{b}\\
\tilde{\boldsymbol{\omega}} & = & x^{a}\left(\boldsymbol{\lambda}_{a}-b_{ab}\boldsymbol{\kappa}^{b}\right)+\mathbf{d}\left(\ln x^{2}\right)
\end{eqnarray*}
substitution into $\boldsymbol{\Lambda}^{a}$ gives
\begin{eqnarray*}
\boldsymbol{\Lambda}^{a} & \equiv & b^{ab}\tilde{\boldsymbol{\omega}}_{\,\,b}^{c}\wedge\boldsymbol{\lambda}_{c}-b^{cb}\boldsymbol{\lambda}_{b}\wedge\tilde{\boldsymbol{\omega}}_{\,\,c}^{a}-2\tilde{\boldsymbol{\omega}}\wedge b^{ab}\boldsymbol{\lambda}_{b}+\mathbf{d}b^{ab}\boldsymbol{\lambda}_{b}\\
 & = & b^{ab}\left(-2\Delta_{db}^{ce}\left(x^{d}\boldsymbol{\lambda}_{e}-b_{ef}x^{d}\boldsymbol{\kappa}^{f}+\frac{2}{x^{2}}x_{e}\boldsymbol{\kappa}^{d}\right)\right)\wedge\boldsymbol{\lambda}_{c}\\
 &  & -b^{cb}\boldsymbol{\lambda}_{b}\wedge\left(-2\Delta_{dc}^{ae}\left(x^{d}\boldsymbol{\lambda}_{e}-b_{ef}x^{d}\boldsymbol{\kappa}^{f}+\frac{2}{x^{2}}x_{e}\boldsymbol{\kappa}^{d}\right)\right)\\
 &  & -2\left(x^{e}\left(\boldsymbol{\lambda}_{e}-b_{ef}\boldsymbol{\kappa}^{f}\right)+\mathbf{d}\left(\ln x^{2}\right)\right)\wedge b^{ab}\boldsymbol{\lambda}_{b}+\mathbf{d}b^{ab}\boldsymbol{\lambda}_{b}
\end{eqnarray*}
Dropping the irrelevant $\boldsymbol{\kappa}^{a}$ terms (since only
$\boldsymbol{\lambda}_{a}\wedge\boldsymbol{\lambda}_{b}$ terms violate
the involution),
\begin{eqnarray*}
\boldsymbol{\Lambda}^{a} & \cong & b^{ab}\left(-2\Delta_{db}^{ce}x^{d}\boldsymbol{\lambda}_{e}\right)\wedge\boldsymbol{\lambda}_{c}+b^{cb}\boldsymbol{\lambda}_{b}\wedge2\Delta_{dc}^{ae}\left(x^{d}\boldsymbol{\lambda}_{e}\right)-2\left(x^{e}\boldsymbol{\lambda}_{e}+\mathbf{d}\left(\ln x^{2}\right)\right)\wedge b^{ab}\boldsymbol{\lambda}_{b}+\mathbf{d}b^{ab}\boldsymbol{\lambda}_{b}
\end{eqnarray*}
Also, since $b_{ab}$ depends only on $x^{a}$, the derivatives $\mathbf{d}b_{ab}$
depend only on $\boldsymbol{\kappa}^{a}$ and may be dropped,
\begin{eqnarray*}
\boldsymbol{\Lambda}^{a} & \cong & \left(b^{bc}2\Delta_{db}^{ae}x^{d}+b^{ab}2\Delta_{db}^{ce}x^{d}+2x^{e}b^{ac}\right)\boldsymbol{\lambda}_{c}\wedge\boldsymbol{\lambda}_{e}\\
 & = & \left(b^{bc}\left(\delta_{d}^{a}\delta_{b}^{e}-\delta^{ae}\delta_{bd}\right)x^{d}+b^{ab}\left(\delta_{d}^{c}\delta_{b}^{e}-\delta^{ce}\delta_{bd}\right)x^{d}+2x^{e}b^{ac}\right)\boldsymbol{\lambda}_{c}\wedge\boldsymbol{\lambda}_{e}\\
 & = & \left(b^{ac}x^{e}-x_{b}b^{bc}\delta^{ae}\right)\boldsymbol{\lambda}_{c}\wedge\boldsymbol{\lambda}_{e}\\
 & = & \left(2x^{a}x^{c}x^{e}-x^{2}\left(\delta^{ac}x^{e}+\delta^{ae}x^{c}\right)\right)\boldsymbol{\lambda}_{c}\wedge\boldsymbol{\lambda}_{e}\\
 & = & 0
\end{eqnarray*}
Therefore, $\boldsymbol{\kappa}^{a}$ is in involution for this solution.

For the involution of $\boldsymbol{\lambda}^{a}$ we require $\boldsymbol{\Sigma}^{a}$
to be linear in $\boldsymbol{\lambda}^{a}$. In fact, it vanishes
identically:
\begin{eqnarray*}
\boldsymbol{\Sigma}^{a} & \equiv & b^{ae}\tilde{\boldsymbol{\omega}}_{\,\,e}^{c}\wedge b_{cb}\boldsymbol{\kappa}^{b}+\boldsymbol{\kappa}^{a}\wedge\tilde{\boldsymbol{\omega}}-b^{ac}\mathbf{d}b_{cb}\boldsymbol{\kappa}^{b}-\boldsymbol{\kappa}^{c}\wedge\tilde{\boldsymbol{\omega}}_{\,\,c}^{a}-\tilde{\boldsymbol{\omega}}\wedge\boldsymbol{\kappa}^{a}\\
 & = & b^{ae}\left(-2\Delta_{de}^{cg}\left(-b_{gf}x^{d}\boldsymbol{\kappa}^{f}+\frac{2}{x^{2}}x_{g}\boldsymbol{\kappa}^{d}\right)\right)\wedge b_{cb}\boldsymbol{\kappa}^{b}\\
 &  & -\boldsymbol{\kappa}^{c}\wedge\left(-2\Delta_{dc}^{ab}\left(-b_{bf}x^{d}\boldsymbol{\kappa}^{f}+\frac{2}{x^{2}}x_{b}\boldsymbol{\kappa}^{d}\right)\right)\\
 &  & +2\boldsymbol{\kappa}^{a}\wedge\left(x^{a}\left(-b_{ab}\boldsymbol{\kappa}^{b}\right)+\mathbf{d}\left(\ln x^{2}\right)\right)-b^{ac}\mathbf{d}b_{cb}\boldsymbol{\kappa}^{b}\\
 & = & \frac{1}{\left(x^{2}\right)^{2}}\left(x^{2}\delta_{f}^{a}x_{b}-x^{a}\left(4x_{f}x_{b}-2x_{b}x_{f}-2x_{b}x_{f}+x^{2}\delta_{bf}\right)+4x^{a}x_{b}x_{f}-2x^{2}\delta_{f}^{a}x_{b}\right)\boldsymbol{\kappa}^{f}\wedge\boldsymbol{\kappa}^{b}\\
 &  & +\frac{1}{\left(x^{2}\right)^{2}}\left(2x^{a}x_{c}x_{f}-\delta_{f}^{a}x_{c}x^{2}\right)\boldsymbol{\kappa}^{c}\wedge\boldsymbol{\kappa}^{f}+\frac{2}{x^{2}}x_{c}\boldsymbol{\kappa}^{c}\wedge\boldsymbol{\kappa}^{a}\\
 &  & -2\frac{1}{\left(x^{2}\right)^{2}}x_{b}x^{2}\boldsymbol{\kappa}^{a}\wedge\boldsymbol{\kappa}^{b}+\frac{2}{x^{2}}2x_{c}\boldsymbol{\kappa}^{a}\wedge\mathbf{d}x^{c}\\
 &  & +\frac{1}{\left(x^{2}\right)^{2}}\left(2x^{2}x_{b}\delta_{c}^{a}+2x^{a}2x_{b}x_{c}-2x^{a}\delta_{bc}x^{2}-4x_{b}x_{c}x^{a}+2\delta_{b}^{a}x_{c}x^{2}\right)\mathbf{d}x^{c}\boldsymbol{\kappa}^{b}\\
 & = & -\frac{1}{x^{2}}\delta_{f}^{a}x_{c}\boldsymbol{\kappa}^{f}\wedge\boldsymbol{\kappa}^{c}+\frac{1}{x^{2}}\delta_{f}^{a}x_{c}\boldsymbol{\kappa}^{f}\wedge\boldsymbol{\kappa}^{c}+\frac{2}{x^{2}}x_{c}\boldsymbol{\kappa}^{c}\wedge\boldsymbol{\kappa}^{a}+\frac{2}{x^{2}}2x_{c}\boldsymbol{\kappa}^{a}\wedge\mathbf{d}x^{c}-2\frac{1}{\left(x^{2}\right)^{2}}x_{b}x^{2}\boldsymbol{\kappa}^{a}\wedge\boldsymbol{\kappa}^{b}\\
 &  & +\frac{1}{\left(x^{2}\right)^{2}}\left(2x^{2}x_{b}\delta_{c}^{a}+2x^{a}2x_{b}x_{c}-2x^{a}\delta_{bc}x^{2}-4x_{b}x_{c}x^{a}+2\delta_{b}^{a}x_{c}x^{2}\right)\mathbf{d}x^{c}\boldsymbol{\kappa}^{b}\\
 & = & \frac{1}{x^{2}}\left(2\delta_{c}^{a}x_{b}+2\delta_{b}^{a}x_{c}\right)\boldsymbol{\kappa}^{c}\boldsymbol{\kappa}^{b}\\
 & = & 0
\end{eqnarray*}
where we use $\mathbf{d}x^{b}=\boldsymbol{\kappa}^{b}$ in the last
step. 

The description of the spacetime and momentum submanifolds follow
from the solution,
\begin{eqnarray*}
\tilde{\boldsymbol{\omega}}_{\,\,b}^{a} & = & -2\Delta_{db}^{ac}\left(x^{d}\mathbf{d}u_{c}-b_{cf}x^{d}\mathbf{d}x^{f}+\frac{2}{x^{2}}x_{c}\mathbf{d}x^{d}\right)\\
\tilde{\boldsymbol{\omega}}^{a} & = & \frac{1}{2}\left(\mathbf{d}x^{a}-b^{ab}\mathbf{d}u_{b}\right)\\
\tilde{\boldsymbol{\omega}}_{a} & = & \mathbf{d}u_{a}+b_{ab}\mathbf{d}x^{b}\\
\tilde{\boldsymbol{\omega}} & = & x^{a}\left(\mathbf{d}u_{a}-b_{ab}\mathbf{d}x^{b}\right)+\mathbf{d}\left(\ln x^{2}\right)
\end{eqnarray*}
When we hold $u_{a}$ constant
\begin{eqnarray*}
\tilde{\boldsymbol{\omega}}_{\,\,b}^{a} & = & 2\Delta_{db}^{ac}\left(b_{cf}x^{d}\mathbf{d}x^{f}-\frac{2}{x^{2}}x_{c}\mathbf{d}x^{d}\right)\\
\tilde{\boldsymbol{\omega}}^{a} & = & \frac{1}{2}\mathbf{d}x^{a}\\
\tilde{\boldsymbol{\omega}} & = & \frac{1}{2}\mathbf{d}\left(\ln x^{2}\right)
\end{eqnarray*}
with $\tilde{\boldsymbol{\omega}}_{a}=b_{ab}\mathbf{d}x^{b}$. The
submanifold is clearly Lagrangian, the connection $SO\left(4\right)$,
and inner product of the basis forms is Lorentzian,
\begin{eqnarray*}
\left\langle \tilde{\boldsymbol{\omega}}^{a},\tilde{\boldsymbol{\omega}}^{b}\right\rangle  & = & \frac{1}{4}\left\langle \mathbf{d}x^{a},\mathbf{d}x^{b}\right\rangle \\
 & = & \frac{1}{4}b^{ab}
\end{eqnarray*}

On the momentum submanifold, with $x^{a}=x_{0}^{a}$ constant,
\begin{eqnarray*}
\tilde{\boldsymbol{\omega}}_{\,\,b}^{a} & = & -2\Delta_{db}^{ac}x_{0}^{d}\mathbf{d}u_{c}\\
\tilde{\boldsymbol{\omega}}_{a} & = & \mathbf{d}u_{a}\\
\tilde{\boldsymbol{\omega}} & = & x_{0}^{a}\mathbf{d}u_{a}
\end{eqnarray*}
where $\tilde{\boldsymbol{\omega}}^{a}=-\frac{1}{2}b^{ab}\left(x_{0}^{\alpha}\right)\mathbf{d}u_{b}$.
This is again Lagrangian with inner product
\begin{eqnarray*}
\left\langle \tilde{\boldsymbol{\omega}}_{a},\tilde{\boldsymbol{\omega}}_{b}\right\rangle  & = & \left\langle \mathbf{d}u_{a},\mathbf{d}u_{b}\right\rangle \\
 & = & b_{ab}\left(x_{0}^{\alpha}\right)
\end{eqnarray*}
Since the metric is now constant, the momentum submanifold is flat.
The Yang-Mills field strength has the form
\begin{eqnarray*}
\mathbf{F}_{\quad b}^{a} & = & \mathbf{d}\tilde{\boldsymbol{\omega}}_{\,\,b}^{a}-\tilde{\boldsymbol{\omega}}_{\,\,b}^{c}\wedge\tilde{\boldsymbol{\omega}}_{\,\,c}^{a}\\
 & = & -4\Delta_{db}^{ec}\Delta_{ge}^{af}x_{0}^{d}x_{0}^{g}\mathbf{d}u_{c}\wedge\mathbf{d}u_{f}\\
 & = & -\Delta_{bd}^{ac}b_{0}^{df}\mathbf{d}u_{c}\wedge\mathbf{d}u_{f}
\end{eqnarray*}
which is of the form of a pure rescaling and can therefore, be made
to vanish by a conformal transformation, $\phi=-x_{0}^{\alpha}u_{\alpha}$.
\end{document}